\documentclass[12pt]{article}
\usepackage{epsf,amsfonts,amssymb,epsfig,amsmath}
\renewcommand{\text}[1]{#1}

\newcommand{\be}{\begin{equation}}
\newcommand{\ee}{\end{equation}}
\newcommand{\ben}{\begin{displaymath}}
\newcommand{\een}{\end{displaymath}}
\newcommand{\bea}{\begin{eqnarray}}
\newcommand{\eea}{\end{eqnarray}}
\newcommand{\bean}{\begin{eqnarray*}}
\newcommand{\eean}{\end{eqnarray*}}
\newcommand{\nn}{\nonumber \\}
\newcommand{\ba}{\begin{array}}
\newcommand{\ea}{\end{array}}
\newcommand{\bi}{\begin{itemize}}
\newcommand{\ei}{\end{itemize}}

\newcommand{\reef}[1]{(\ref{#1})}

\def\e{\epsilon}
\def\ve{\varepsilon}

\def\e{\epsilon}

\newcommand{\bbR}{{\mathbb{R}}}

\newcommand{\sla}{\slash\!\!\!\!}

\newcommand{\dslash}{\slash\!\!\!\partial}

\begin{document}

\makeatletter
\renewcommand{\theequation}{\thesection.\arabic{equation}}
\@addtoreset{equation}{section}
\makeatother

\baselineskip 18pt

\begin{titlepage}

\vfill

\begin{flushright}
Imperial/TP/2009/JG/06\\
DESY 09-107\\
\end{flushright}

\vfill

\begin{center}
   \baselineskip=16pt
 {\Large\bf  Schr\"odinger invariant solutions of type IIB\\ 
 with enhanced supersymmetry}
  \vskip 1.5cm
      Aristomenis Donos$^1$ and Jerome P. Gauntlett$^2$\\
   \vskip .6cm
      \begin{small}
      $^1$\textit{DESY Theory Group, DESY Hamburg\\
        Notkestrasse 85, D 22603 Hamburg, Germany}
        \end{small}\\*[.6cm]
        \begin{small}
      $^2$\textit{Theoretical Physics Group, Blackett Laboratory, \\
        Imperial College, London SW7 2AZ, U.K.}
        \end{small}\\*[.6cm]
      \begin{small}
      $^2$\textit{The Institute for Mathematical Sciences, \\
        Imperial College, London SW7 2PE, U.K.}
        \end{small}\\*[.6cm]
   \end{center}

\vfill

\begin{center}
\textbf{Abstract}
\end{center}

\begin{quote}
We construct the Killing spinors for a class of supersymmetric solutions of type IIB supergravity
that are invariant under the non-relativistic Schr\"odinger algebra. The solutions depend on a
five-dimensional Sasaki-Einstein space and it has been shown that they admit two Killing spinors.
Here we will show that, for generic Sasaki-Einstein space, there are special subclasses of solutions which admit six Killing spinors
and we determine the corresponding superisometry algebra. We also show that for the special case that the Sasaki-Einstein space is the
round five-sphere, the number of Killing spinors can be increased to twelve.
\end{quote}

\vfill

\end{titlepage}
\setcounter{equation}{0}

\section{Introduction}

Consider the class of type IIB supergravity solutions of \cite{dg2} given by
\begin{align}\label{solution}
ds^{2} & = \Phi^{-{1}/{2}}\left[2\, dx^{+}dx^{-}+h\,\left(dx^{+}\right)^{2}+dx_{2}^{2}+dx_{3}^{2}\right]+
 \Phi^{{1}/{2}}\, ds^{2}\left(CY_{3}\right)\nn
F & =\left(1+\ast_{10}\right)dx^{+}\wedge dx^{-}\wedge dx_{2}\wedge dx_{3}\wedge d \Phi^{-1}\nn
G & =dx^{+}\wedge W
\end{align}
where $F$ is the self-dual five-form, $G$ is the complex three-form and the dilaton and axion vanish.
Here $\Phi$ and $h$ are scalars and $W$ is a complex two-form defined on the Calabi-Yau three-fold $CY_3$ that
satisfy
\bea\label{eqsom}
\nabla^{2}_{CY}\Phi&=&0\nn
\nabla^{2}_{CY}h+|W|^2_{CY} & =&0\nn
dW=d\ast_{CY}W & =&0
\eea
where $|W|^2_{CY}\equiv\tfrac{1}{2}W_{mn}W^{*mn}$ with indices raised with respect to the $CY_3$ metric.

Our focus will be on cases where the $CY_{3}$ is a cone over a five dimensional Sasaki-Einstein space $SE_{5}$
and the harmonic function $\Phi$ has a source at the apex of the cone:
\bea\label{cone}
ds^{2}\left(CY_{3}\right)&=&dr^{2}+r^{2}\, ds^{2}\left(SE_{5}\right)\nn
\Phi&=&r^{-4}~.
\eea
When $W=h=0$ this gives the well known $AdS_5\times SE_5$ class of solutions describing D3-branes sitting at the
apex of the cone. We will be most interested in the case where the deformation $W$ is
given by
\bea\label{phiandw}
W&=&d\left(r^{2}\sigma\right)
\eea
where $\sigma$ is a complex one-form dual to a Killing vector on $SE_{5}$, and $h$ is obtained by solving
the second equation in \reef{eqsom}. This particular class of solutions, for the special case of $W$ being
real, was also independently discovered in \cite{pilch} using a solution generating technique.

An interesting feature of this class of solutions is that
they are invariant under the Schr\"odinger algebra \cite{dg2}\cite{pilch}. The current interest in these solutions
is that they might provide a good holographic description of non-relativistic systems that are invariant
under such symmetry \cite{Son:2008ye}\cite{Balasubramanian:2008dm}. The principal aim of this paper is to carry out
a careful study of the supersymmetry preserved by these solutions, building on the observations of \cite{dg2}\cite{pilch}.
In particular, we will see that for special subclasses of solutions fixed by ($h$, $\sigma$), there is the possibility of extra
``supernumerary'' Killing spinors. In many ways, the analysis is reminiscent of the supersymmetry enhancement that occurs for
plane wave solutions \cite{Cvetic:2002si}\cite{Gauntlett:2002cs}.
Note that in addition to the papers \cite{dg2}\cite{pilch} supersymmetric solutions of type IIB or $D=11$ supergravity with Schr\"odinger or
Schr\"odinger$(z)$ symmetry, where $z$ is the dynamical exponent, have also been discussed in \cite{Hartnoll:2008rs}-\cite{Ooguri:2009cv}.

For a general $CY_3$ (i.e not necessarily a cone) and with $W=h=0$ the solutions \reef{solution}, \reef{eqsom} describe
D3-branes transverse to the $CY_3$ and preserve, generically, four ``Poincar\'e" supersymmetries satisfying
\be
\gamma_{D3}\ve =\ve, \quad \nabla^{CY}\ve=0
\ee
where we have defined the D3-brane projection
\be
\gamma_{D3}\equiv i\Gamma^{+-23}~.
\ee
Here we are using a light cone frame and $\nabla^{CY}$ is the Levi-Civita connection on
$\bbR^{1,3}\times CY_3$.
As shown in \cite{dg2} when the two-form $W$ on $CY_3$ is primitive and with no $(0,2)$ component
(i.e. just $(1,1)$ and/or $(2,0)$ components)
then two of these Killing spinors, satisfying the additional projection
$\Gamma^+\epsilon=0$, are preserved, and furthermore the functional form of the Killing spinors are the same as
those for the
$W=h=0$ solutions.

When the $CY_3$ is a cone, as in \reef{cone}, and $W=h=0$ the solutions are $AdS_5\times SE_5$ solutions and, generically, in addition to the four
Poincar\'e supersymmetries there are an extra four ``superconformal" supersymmetries. Recall that if one takes the Lie derivative
of the Poincar\'e Killing spinors with respect to the special conformal Killing vectors, one
obtains the special conformal Killing spinors. Here we will study the possibility of an analogous enhancement
of supersymmetry when $W\ne 0$. We will focus on the case when $W$ is of the form
given in \reef{phiandw} when we know the solutions are invariant under the Schr\"odinger algebra.
In particular, these solutions posses a Killing vector corresponding to
special conformal transformations and naively one might think that after taking the Lie-derivative of
the two Poincar\'e supersymmetries satisfying $\Gamma^+\varepsilon=0$ mentioned above,
one would obtain new superconformal supersymmetries. We will show that this is
in fact {\it not} the case and that the Lie derivative vanishes. However, we shall see that for special choices of ($h$,$\sigma$) there
can be two additional Poincar\'e supersymmetries, with $\Gamma^+\ve\ne0$ and whose functional
form depends on $W$, and that the Lie derivative of these give rise to two additional superconformal supersymmetries.
Generically, then, one has six supersymmetries\footnote{Note that solutions of $D=11$ supergravity with Schr\"odinger symmetry and six Killing spinors were also found
in \cite{Ooguri:2009cv}.}
which can be viewed as deformations of four Poincar\'e and two special conformal supersymmetries.
We will also show that for the special case when the $SE_5$ is a round $S^5$ the supersymmetry can be enhanced to
eight Poincar\'e and four special conformal supersymmetries.

Having constructed the explicit Killing spinors we can use them to study the superisometry algebra using the technique of
\cite{Gauntlett:1998kc}\cite{FigueroaO'Farrill:1999va}.
In particular, the Grassmann odd-odd part of this algebra is obtained by constructing Killing vectors as bi-linears in the Killing spinors.
The odd-even part of the algebra is obtained by taking the Lie derivative of the Killing spinors with respect to the Killing vectors.
The resulting super-Schr\"odinger algebras that we obtain are consistent with those\footnote{For other work on super-Schr\"odinger algebras
see \cite{Gauntlett:1990xq}-\cite{Nakayama:2009cz}.}
 found in \cite{Sakaguchi:2008ku}.
The two Poincar\'e supersymmetries found in \cite{dg2} are ``kinematical'' supersymmetries, with anti-commutator giving the central number operator
of the Schr\"odinger algebra. The two new Poincar\'e Killing spinors
that we find here are ``dynamical'' supersymmetries, with anti-commutator giving the non-relativisitic Hamiltonian $H$ and they lead to a positive spectrum for $H$.

We will also briefly consider the more general class of solutions \reef{solution}, \reef{eqsom}, \reef{cone}
when $W$ is of the form $W=d(r^z \sigma)$ for $z>2$.
This class of solutions has Schr\"odinger$(z)$ symmetry, where $z$ is the dynamical exponent.
We shall find while there cannot be any superconformal supersymmetries, for special subclasses of
solutions it is possible
to have dynamical supersymmetries in addition to the kinematical supersymmetries found in \cite{dg2}.

The plan of the rest of the paper is as follows. In section 2 we analyse in detail the conditions for supersymmetry.
We have summarised some of the calculations in section 2.4 where we also present some explicit examples. Section 3
studies the superisometry algebra and section 4 briefly concludes. Appendix A contains some useful results about
$CY_3$ cones, appendix B a technical derivation arising in section 2, and appendix C a brief discussion of the $z>2$ solutions.

\section{Construction of Killing spinors}
We will now carry out our analysis of the Killing spinor equations for the class of solutions
given above in \reef{solution} and \reef{cone}. We will focus on the case when $W=d(r^z\sigma)$ with $z=2$,
reserving some comments about the case when $z>2$ to the appendix.
Our main results are summarised in section 2.4.

The conditions for these solutions to admit IIB Killing spinors $\varepsilon$ are given by
\bea
D_{M}\varepsilon+\tfrac{i}{16}{\sla{F}}{\Gamma}_{M}\varepsilon+\tfrac{1}{16}\left(\Gamma_M\sla{G}+2~{\sla{G}}{\Gamma}_{M}\right)\varepsilon^{\ast}&=&0\label{ks1}\\
{\sla{G}}\varepsilon&=&0\label{ks2}
\eea
where e.g. $\sla{F}\equiv\tfrac{1}{5!}\Gamma^{P_1\dots P_5}\Gamma_{P_1\dots P_5}$.
We will use the orthonormal frame given by
$e^+=\Phi^{-1/4}dx^+$, $e^-=\Phi^{-1/4}(dx^-+\tfrac{1}{2}hdx^+)$, $e^2=\Phi^{-1/4}dx^2$, $e^3=\Phi^{-1/4}dx^3$, $e^a=\Phi^{1/4}f^a$,
where $f^a$ is an orthonormal frame for the $CY_3$ metric: $f^a f^a=ds^2(CY_3)$.
The gamma-matrices with $D=10$ tangent space indices,
$\Gamma^A=\{\Gamma^+$, $\Gamma^-$, $\Gamma^2$, $\Gamma^3$, $\Gamma^a\}$,
satisfy $\{\Gamma^A,\Gamma^B\}=2\eta^{AB}$ (with e.g. $\eta_{\pm}=+1$) and indices can be raised and lowered using the tangent space metric $\eta$.
We are using the conventions of type IIB supergravity given in \cite{Gauntlett:2005ww} and in particular
$\Gamma_{11}\varepsilon  =-\varepsilon$ where $\Gamma_{11}\equiv\Gamma_{+-23456789}$. We find it convenient to work
in basis in which the 
gamma-matrices are real and $\varepsilon^c=\varepsilon^*$.

It will be helpful to introduce some further notation. We let $x^m$ be coordinates on the $CY_3$, and we will write
$x^m=(r,x^\mu)$ where $x^{\mu}$ are coordinates on $SE_5$. Similarly for the orthonormal frame on the $CY_3$ cone
we write $f^a=(dr,r\bar f^\alpha)$ where $\bar f^{\alpha}$ is an orthonormal frame for the $SE_5$ metric: $\bar f^\alpha \bar f^\alpha =ds^2(SE_5)$.
Correspondingly we also write $\Gamma^a=(\Gamma^r,\Gamma^\alpha)$ (with e.g. $(\Gamma^r)^2=(\Gamma^\alpha)^2=1)$. We emphasise that for forms defined on
the $CY_3$ space tangent space indices will always refer to the frame $f^a$ so, for example, $d\Phi=(d\Phi)_af^a$. Furthermore for such forms we
use the slash notation to mean e.g.
\be
\sla{W}\equiv \tfrac{1}{2}W_{ab}\Gamma^{ab},\qquad \dslash \Phi \equiv (d\Phi)_a \Gamma^a
\ee
(note that this differs by a factor of $\Phi^{1/4}$ from the slash notation used for the ten-dimensional fields in \reef{ks1}, \reef{ks2}.)
Similarly tangent frame indices on $\sigma$ and its derivatives will refer to the frame $\bar f^{\alpha}$ e.g. $\sigma=\sigma_\alpha \bar f^\alpha$.
Unless otherwise stated, all gamma-matrices will be understood to be tangent space gamma-matrices $\Gamma^A$ satisfying $\{\Gamma^A,\Gamma^B\}=\eta^{AB}$.

\subsection{Analysis for $M=-,2,3$}
Let us consider the Killing spinor equation \reef{ks1} when the coordinate index $M=-,2,3$. It will be convenient
to define $X^I=(x^-, x^i)$, with $i=2,3$. For these coordinates, the Killing spinor equation \reef{ks1} takes the form
\be\label{ieq}
\partial_{I}\varepsilon=\frac{r}{2}\left(1-\gamma_{D3}\right)\Gamma^r\Gamma_I\varepsilon -\frac{r^2}{16}\Gamma^+\sla{W}\Gamma_I\epsilon^*~.
\ee
This easily gives
\be
\partial_I\partial_{J}\varepsilon=\frac{r^3}{32}\Gamma_I\Gamma_J\{(1-\gamma_{D3})\Gamma^r,\Gamma^+\sla{W}\}\varepsilon^*
\ee
and hence, after anti-symmetrising on $I$ and $J$,
\bea
\{(1-\gamma_{D3})\Gamma^r,\Gamma^+\sla{W}\}\varepsilon^*=0\nn
\partial_I\partial_{J}\varepsilon=0~.
\eea
We thus can write $\varepsilon=\varepsilon_{0}+x^{I}\varepsilon_{I}$ with $\varepsilon_{0}$ and $\varepsilon_{I}$ depending on $x^{+}$ and the $CY_{3}$ coordinates $x^m$.
Substituting back into \reef{ieq} we obtain the unique solution
\begin{equation}\label{bop}
\varepsilon=\varepsilon_{0}+\frac{r}{2}\Gamma^r (x^I\Gamma_I)\left(1+\gamma_{D3}\right)\varepsilon_{0} -\frac{r^2}{16}\Gamma^+\sla{W}(x^I\Gamma_I)\varepsilon_0^*~.
\end{equation}
We next decompose $\varepsilon_0$ into eigenvalues of $\gamma_{D3}$. It will turn out to be convenient to do this in the following way:
\be\label{blip}
\varepsilon_{0}  = r^{-1/2}\Gamma^r\epsilon_++r^{1/2}\epsilon_-
\ee
where
\be
\gamma_{D3}\epsilon_{\pm}  =\pm\epsilon_{\pm}
\ee
and $\epsilon_{\pm}$ depend only on the coordinates $x^{+},x^m$.

It is now helpful to substitute \reef{bop},\reef{blip} into \reef{ks2}. The terms that are dependent and independent
of the coordinates $x^I$ must each separately vanish and after projecting with $(1/2)(1\pm \gamma_{D3})$ we deduce that
\begin{align}
\Gamma^+\sla{W}\epsilon_{\pm} & =0\label{projpm}\\
\Gamma^+\sla{W}\epsilon_{\pm}^* & =0\label{projpms}\\
\Gamma^+\sla{W}\Gamma^r\epsilon_{+} & =0\label{projr}
\end{align}
and that the ten-dimensional Killing spinor can be written as
\begin{equation}\label{tds}
\varepsilon=r^{1/2}\epsilon_{-}+\left[r^{-1/2}\Gamma^r-r^{1/2} (x^I\Gamma_I)\right]\epsilon_{+}
-\frac{r^{3/2}}{16}\Gamma^+\sla{W}(x^I\Gamma_I)\Gamma^r\epsilon_+^*~.
\end{equation}

Observe that with $W=d(r^z\sigma)$ \reef{projpm}, \reef{projr} imply for any $z$ that
\begin{equation}\label{sigone}
\Gamma^{+}({\sigma}_\alpha\Gamma^\alpha)\epsilon_{+}=0~.
\end{equation}

\subsection{Analysis for $M=m$}
We next consider \reef{ks1} for $M=m$. It is useful to define
\be
\nabla^{CY}_m\varepsilon = (\partial_m+\frac{1}{4}\omega_{mab}^{CY}\Gamma^{ab})\varepsilon~.
\ee
where $\omega^{CY}$ is the spin connection on the $CY_3$ with respect to the frame $f^a$.
We next note that since the $CY_3$ is a cone we have
\be\label{lem1}
\nabla^{CY}_m(2r\Gamma^r)
=\nabla^{CY}_m(\dslash r^2)=
2\Gamma_a f^a_m
\ee
where $f^a\equiv f^a_m dx^m$.
After separately considering the $x^I$ dependent and independent components and projecting with $(1/2)(1\pm \gamma_{D3})$, we find that the
$M=m$ component of \reef{ks1} gives rise to three equations
\begin{align}
\nabla^{CY}_{m}\epsilon_{+}+\frac{1}{16}\Gamma^+\sla{W}(\Gamma_{a}f^a_m)\e^*_+-\frac{r}{16}\Gamma^+\nabla^{CY}_m\sla{W}\Gamma^r\e^*_+ & =0\label{covp1}\\
\nabla^{CY}_{m}\epsilon_{+}+\frac{1}{8}\Gamma^r\Gamma^+\sla{W}(\Gamma_{a}f^a_m)\Gamma^r\e^*_+ + \frac{1}{16}\Gamma^r (\Gamma_{a}f^a_m)\Gamma^+\sla{W}\Gamma^r\e^*_+ &=0\label{covp2}\\
\nabla^{CY}_{m}\epsilon_{-}+\frac{1}{8}\Gamma^{+}\sla{W}(\Gamma_{a}f^a_m)\epsilon_{-}^{\ast}& =0~.\label{eq:grav_IIB}
\end{align}
Note that these imply
\be
\Gamma^+\nabla^{CY}_m\e_\pm=0
\ee
and then using \reef{projpm}-\reef{projr} we get
\be\label{pcds}
\Gamma^+\nabla_m^{CY}\sla{W}\e_\pm=\Gamma^+\nabla_m^{CY}\sla{W}\e_\pm^*=0~.
\ee

Next, using the fact that for $z=2$ we have
\be\label{z2l1}
[\nabla^{CY}_m\sla{W},\Gamma^r]=2\nabla^{CY}_m W_{ar}\Gamma^a=0,
\ee
where in the last step we used \reef{z2wd}, and combining with \reef{pcds} we deduce that
the last term in \reef{covp1} vanishes. Therefore, we can solve \reef{covp1} by writing
\bea
\epsilon_{+} &=&\psi_{+}-\frac{r}{16}\Gamma^+\sla{W}\Gamma^r  \psi_{+}^{\ast}\label{defepplus}\\
\nabla^{CY}_m\psi_+ &=&0
\eea
with $\psi_+$ satisfying $\gamma_{D3}\psi_+=i\psi_+$ and constraints arising from \reef{projpm}-\reef{projr}:
\begin{align}
\Gamma^+\sla{W}\psi_{+}=\Gamma^+\sla{W}\psi_{+}^*=
\Gamma^+\sla{W}\Gamma^r\psi_{+} & =0~.\label{psiproj}
\end{align}
Note that we could solve \reef{eq:grav_IIB} in a similar way, but we delay doing that for a moment.

The compatibility of \reef{covp1} and \reef{covp2} imply that
\be
\Gamma^+\left[-\sla{W}\Gamma_\alpha-2\Gamma^r\sla{W}\Gamma_\alpha\Gamma^r+\Gamma^r\Gamma_\alpha\sla{W}\Gamma^r\right]\psi_+^*=0
\ee
which implies that
\be
\Gamma^+\left[\nabla^{SE}_\alpha\sigma_\beta\Gamma^\beta+2\sigma_\alpha\Gamma^r-\sigma_\beta\Gamma_\alpha{}^\beta\Gamma^r\right]\psi^*_+=0~. \label{consistency_cond}
\ee

\subsection{Analysis for $M=+$}
We now consider \reef{ks1} for $M=+$. We find
\bea
\partial_+\varepsilon+\frac{r}{2}\Gamma^{-r}(1+\gamma_{D3})\varepsilon +\frac{hr}{4}\Gamma^{+r}(1+\gamma_{D3})\varepsilon
+\frac{r^2}{4}\Gamma^{+}\dslash h\varepsilon\nn+\frac{r^2}{8}\Gamma^+\Gamma^-\sla{W}\varepsilon^*
+\frac{r^2}{16}\Gamma^-\Gamma^+\sla{W}\ve^*=0~.
\eea
After substituting in the expression for $\varepsilon$ given in \reef{tds}, isolating the terms depending on $x^I$ and then projecting with
$(1/2)(1\pm \gamma_{D3})$
we are led to
\begin{align}
\partial_{+}\epsilon_{+} & =0\label{benasq}\\
\partial_+\epsilon_-+\Gamma^-\epsilon_+ +  \frac{r}{4}\sla{W}\Gamma^r\e^*_+ +        \Gamma^+\left(\frac{r}{4}\dslash h \Gamma^r+\frac{h}{2}\right)\e_+
-\frac{r}{16}\Gamma^-\Gamma^+\sla{W}\Gamma^r\epsilon_+^*&=0\label{thirdeq}\\
\Gamma^+\dslash h\epsilon_+   +\sla{W}\e_+^*            -\frac{r}{16}\Gamma^+\sla{W}\sla{W}^*\Gamma^r\epsilon_+ &=0\label{tthy}    \\
\Gamma^+\dslash h\epsilon_- +  \sla{W}\epsilon_-^*&=0~.\label{fourtheqn}
\end{align}

We would now like to argue that $\Gamma^+\psi_+=0$. We start by
substituting \reef{defepplus} into \reef{sigone} to obtain
\be
\Gamma^{+}({\sigma}_\alpha\Gamma^\alpha)\psi_{+}=0~.
\ee
Differentiating this and using $\nabla^{CY}_m\psi_+=0$
we obtain
\be
\Gamma^+\left[\sigma_\alpha\Gamma^r-\nabla_\alpha^{SE}\sigma_\beta \Gamma^\beta\right]\psi_+=0
\ee
(one can use \reef{diffconerw} to obtain this).
After contracting with $\sigma^{*\alpha}$ we get
\be
\Gamma^+\left[|\sigma|^2\Gamma^r-\sigma^{*\alpha}\nabla_\alpha^{SE}\sigma_\beta \Gamma^\beta\right]\psi_+=0~.
\label{fstcond}
\ee
We next substitute \reef{defepplus} into \reef{tthy}
to get
\be
\Gamma^+\left[\dslash{h}-\frac{r}{8} \sla{W}\sla{W^*}\Gamma^r\right]\psi_++\sla{W}\psi^*_+=0~.
\ee
From \reef{eqsom} we deduce that the two terms have different scalings with respect to $r$ and hence
must separately vanish
\bea
\Gamma^+\left[\dslash{h}-\frac{r}{8} \sla{W}\sla{W^*}\Gamma^r\right]\psi_+&=&0\nn
\sla{W^*}\psi_+&=&0~.
\eea
Next using also that $\Gamma^+\sla{W}\psi_+=0$ the first equation implies that
\be
\Gamma^+\left[\dslash{h}+2r\left(|\sigma|^2\Gamma^r+\sigma^{*\alpha}\nabla^{SE}_\beta\sigma_\alpha \Gamma^\beta\right)
\right]\psi_+=0
\ee
and after using \reef{fstcond} we deduce that
\be
\Gamma^+\dslash{h}\psi_+=0
\ee
and hence that
\be
\Gamma^+\psi_+=0~.
\ee

Using this result, we find that \reef{benasq}-\reef{fourtheqn} simplify considerably.
After substituting \reef{defepplus} we now find that
\begin{align}
\partial_{+}\psi_{+} & =0\\
\partial_+\epsilon_-+\Gamma^-\psi_+ +  \frac{r}{4}\sla{W}\Gamma^r\psi^*_+  &=0   \label{two2} \\
\sla{W}\psi^*_+     &=0    \\
\Gamma^+\dslash h\epsilon_- +  \sla{W}\epsilon_-^*&=0~.\label{four2}
\end{align}
We solve \reef{two2} as
\be
\e_-=\psi_--x^+\left(\Gamma^-\psi_++\frac{r}{4}\sla{W}\Gamma^r\psi^*_+\right) \label{em}
\ee
where $\psi_-$ is independent of $x^+$. Compatibility with
\reef{eq:grav_IIB}, and using \reef{z2l1} (for $z=2$),
then implies
\be\label{divpsim}
\nabla^{CY}_m\psi_-+\frac{1}{8}\Gamma^+\sla{W}\Gamma_m\psi^*_-=0~.
\ee

From \reef{projpm} we also deduce that
\be
\sla{W}\psi_+=0~.
\ee
Returning now to \reef{four2} we find that
\bea
\left(\dslash h-\frac{r}{8}\sla{W}\sla{W}^*\Gamma^r\right)\psi_+&=&0\\
\Gamma^+\dslash h \psi_-+\sla{W}\psi_-^*&=&0~.\label{dshpsim}
\eea

Observe that \reef{divpsim} can be solved by taking
\bea
\psi_-=\eta_--\frac{r}{8}\Gamma^+\sla{W}\Gamma^r\eta_-^*
\eea
with
\bea
\nabla^{CY}_m\eta_-= \Gamma^+\sla{W}\eta_-=\Gamma^+\sla{W^*}\eta_-=0~.
\eea
After substituting into \reef{dshpsim}, we obtain
\be
\Gamma^+\left(\dslash h -\frac{r}{8}\sla{W}\sla{W}^*\Gamma^r\right)\eta_-+\sla{W}\eta^*_-=0~.
\ee
After noting from \reef{eqsom} that there are two terms with different scaling behaviours under
scalings of $r$, we deduce that
\bea
\Gamma^+\left(\dslash h -\frac{r}{8}\sla{W}\sla{W}^*\Gamma^r\right)\eta_- &=&0\label{phoo}\\
\sla{W}\eta^*_-&=&0~.
\eea

\subsection{Summary}
We now summarise our analysis so far. For $z=2$ the most general Killing spinor can be written
as a sum of ``Poincar\'e'' and ``superconformal'' Killing spinors\footnote{We have relabelled $\psi_+$ of the last section as $\eta_+$.}:
\be\label{pands}
\varepsilon=\varepsilon_{P}+\varepsilon_{S}
\ee
where
\bea\label{pandsexplicit}
\varepsilon_{P}&= & r^{1/2}\eta_--\tfrac{1}{8}r^{3/2}\Gamma^+\sla{W}\Gamma^r\eta_-^*             \label{pf}\\
\varepsilon_{S}&= & r^{-1/2}\left(\Gamma^{r}-rx^{i}\Gamma_{i}-rx^{+}\Gamma_+ \right)\eta_{+}-\tfrac{1}{4}x^+r^{3/2}\sla{W}\Gamma^{r}\eta_{+}^{\ast}\label{sf}.
\eea
where $x^i=(x^2,x^3)$.
The spinors $\eta_\pm$ only depend on the $CY_3$ coordinates and satisfy the following conditions:
\bea
\nabla_m^{CY}\eta_{-}&=&0\\
\Gamma^+\left(\dslash h -\frac{r}{8}\sla{W}\sla{W}^*\Gamma^r\right)\eta_- &=&0\\
\Gamma^+\sla{W}\eta_-=\sla{W}^*\eta_-&=&0\\
\nabla_m^{CY}\eta_{+}&=&0\\
\left(\dslash h-\frac{r}{8}\sla{W}\sla{W}^*\Gamma^r\right)\eta_+&=&0\label{fone}\\
\sla{W}\eta_+=\sla{W}^*\eta_+&=&0\\
\gamma_{D3}\eta_{\pm}  =\pm \eta_{\pm},\qquad \Gamma^+\eta_+&=&0~.
\eea
In order to get a supersymmetric solution we also need to ensure that the equations of motion \reef{eqsom} are satisfied.
If $W=d(r^2\sigma)$ then $d*_{CY}W=0$ is equivalent to $\sigma$ being a Killing vector
on the $SE_5$ as we discuss in appendix A. Thus we just need to impose $\nabla^{2}_{CY}h+|W|^2_{CY}=0$.

In carrying out further analysis, it is illuminating to make a 4+6 decomposition and write
the ten dimensional Gamma matrices as
\begin{align}\label{tendgam}
\Gamma_{u} & =\tau_{u}\otimes\mathbb{I}_{8\times8},\qquad u=+,-,2,3\\
\Gamma_{a} & =\tau^{D3}\otimes\gamma_{a}
\end{align}
where $\tau^{D3}=i\tau^+\tau^-\tau^2\tau^3$. $\Gamma_{11}=\tau^{D3}\gamma_{(7)}$ where $\gamma_{(7)}=i\gamma^{456789}$ and so
we can write the spinors $\eta_{\pm}$ as
\begin{equation}\label{etpmdecomp}
\eta_{\pm}=q_{\pm}\otimes\zeta_{+}
\end{equation}
with $q_{\pm}$ being constant spinors on
$\mathbb{R}^{1,3}$ such that $\tau^{D3}q_{\pm}=\pm q_{\pm}$, $\tau^+ q_+=0$
and $\zeta_{+}$ a covariantly constant spinor on $CY_{3}$ of positive chirality (see appendix A for more details on our
conventions).

At this stage it is worth pausing to recover the results found in \cite{dg2}. In that paper Killing spinors with $\eta_+=0$ and
$\Gamma^+\eta_-=0$ were considered. As in \cite{dg2}, the above conditions for supersymmetry then reduce to
$\nabla^{CY}_m\eta_-=\sla{W}^*\eta_-=0$. Clearly the former is satisfied with $\eta_-$ as given in
\reef{etpmdecomp}, while the latter condition is satsified if the two-form $W$ on $CY_3$ has no
$(0,2)$ form pieces i.e. it consists of $(1,1)$ and primitive and/or $(2,0)$ two-forms. Note that the functional form of these Killing spinors is
exactly the same as those for $W=0$ and that they comprise two Poincar\'e Killing spinors.
For the special case of the five-sphere,
for a generic $W$ with no $(0,2)$ pieces with respect to one of the complex
structures on $\bbR^6$, there are again just two Poincar\'e Killing spinors that satisfy this condition. However, there
is the possibility of special $W$ that satisfy this condition for other complex structures. In particular,
for $W$ that live in $\bbR^4 \subset \bbR^6$ there can be four Poincar\'e Killing spinors.

We now look for special choices of $W$ and $h$ which give rise to additional Killing spinors.
Given the decomposition \reef{etpmdecomp}, we want to allow $\tau^+ q_-\ne 0$ and so our conditions boil down
to solving the following equations on the $CY_3$ cone
\bea
\left(\dslash h -\frac{r}{8}\sla{W}\sla{W}^*\gamma^r\right)\zeta_+ &=&0\label{theheqn}\\
\sla{W}\zeta_+=\sla{W}^*\zeta_+&=&0\label{dipsy}~.
\eea
Here all gamma-matrics are those on $CY_3$, $\gamma_a$.
The conditions \reef{dipsy} now require that $W=d(r^2\sigma)$ is necessarily of type (1,1) and primitive on the $CY_3$ cone.
Solving \reef{theheqn} for $h$ leads to additional constraints on $W$. Let us summarise the result
(a few more details are presented in appendix B).
Define a one-form $\lambda$ on the $SE_5$ space given by
\be
\lambda\equiv i\mathcal{L}_{\sigma^{\ast}}^{SE_{5}}\sigma
\ee
where the notation means that we are taking the Lie-derivative with respect to the vector field
which is dual, with respect to the $SE_5$ metric, to $\sigma^*$. For $h$ we take
\begin{equation}
h=-r^{2}\left(\left|\sigma\right|_{SE}^{2}+\tfrac{1}{2}(\eta_{SE})^{\mu}\lambda_{\mu}\right),
\end{equation}
where $\eta_{SE}$ is the one-form on $SE_5$ dual to the Reeb Killing vector. It is interesting to observe that
the expression for $h$ is actually negative definite. This can be seen by writing it as
\be
h=-2|S^{(0,1)}|_{CY}^2
\ee
where $S^{(0,1)}_m\equiv (1/2)\left(S_m+i\mathcal{J}_m{}^nS_n\right)$
and
\be
S\equiv r^2\sigma
\ee
is a one-form dual to a Killing vector on $CY_3$.
Finally, we also need to impose that $\nabla^{2}_{CY}h+|W|^2_{CY}=0$.
As we discuss in appendix B this is guaranteed if the two-form
\be
V=dL
\ee
is primitive on the $CY_3$ where we have introduced
\be
L=r^2\lambda
\ee
which is a one-form dual to a Killing vector on $CY_3$.
In appendix B we also show that $V$ is in fact (1,1).

We have shown that these special classes of Schr\"odinger invariant solutions admit Killing spinors of the form \reef{pands}, \reef{pandsexplicit}, \reef{pf}
where the spinors $\eta_\pm$ are functions of the $CY_3$ coordinates $x^m$ only, and satisfy
$\nabla^{CY}_m\eta_\pm=0$, $\gamma_{D3}\eta_{\pm}=\pm \eta_{\pm}$ and $\Gamma^+\eta_+=0$.
For a generic $SE_5$ space, these solutions preserve six supersymmetries, four ``Poincar\'e'' Killing spinors
$\varepsilon_P$ and two ``superconformal'' Killing spinors $\varepsilon_S$.
The number of supersymmetries being preserved is very suggestive that the superisometry algebra is
the ones discussed in \cite{Sakaguchi:2008ku}. In the next section we will confirm this.

For the special case when $SE_5=S^5$, with cone $\bbR^6$, we can get further enhancement of supersymmetry.
In particular, if the two-form $W$ is not generic but is a two-form
on $\bbR^4\in\bbR^6$ then the conditions $\sla{W}\eta_\pm=\sla{W}^*\eta_\pm=0$ that we imposed can be
satisfied by twice as many Killing spinors satisfying $\nabla^{CY}_m\eta_\pm=0$. This leads to preservation of
twelve supersymmetries, eight $\varepsilon_P$ and four $\varepsilon_S$.

We conclude this section by presenting some simple examples for the case of $S^5$.
Explicitly we let $(z_1,z_2,z_3)$ be complex coordinates on $\bbR^6$ and take
\begin{align}\label{wanst}
W&=\left(c_{1}d\bar{z}_{2}+c_{2}\,d\bar{z}_{3}\right)\wedge dz_{1}+\left(c_{3}d\bar{z}_{3}+c_{4}d\bar{z}_{1} \right)\wedge dz_{2} +\left(c_{5}d\bar{z}_{1}+c_{6}d\bar{z}_{2} \right)\wedge dz_{3}
\end{align}
where $c_i$ are complex constants. After writing $W=d(r^2\sigma)$ where $\sigma$ is defined on $S^5$, we find that
\bea
2S\equiv r^2\sigma&=&c_1(\bar z_2 dz_1-z_1d\bar z_2)+c_2(\bar z_3 dz_1-z_1d\bar z_3)+\dots\nn
2S^{(0,1)}&=&-(c_1z_1+c_6 z_3)d \bar z_2-(c_4z_2+c_5 z_3)d\bar z_1-(c_2z_1+c_3z_2)d\bar z_3
\eea
giving
\begin{align}
h&=-\left|c_{1}{z}_{1}+c_{6}\bar{z}_{3}\right|^{2}
-\left|c_{2}\bar{z}_{1}+c_{3}\bar{z}_{2} \right|^{2}
-\left|c_{4}\bar{z}_{2}+c_{5}\bar{z}_{3}\right|^{2}~.
\end{align}
One can directly check that $\nabla^{2}_{CY}h+|W|^2_{CY}=0$ and hence we indeed have
a supersymmetric solution
generically preserving six supersymmetries.
An interesting special case is when $c_{1}=c_{3}=c_{5}\equiv c$
and $c_{2}=c_{4}=c_{6}=0$. We then have
\bea
W&=&c(d\bar z_2\wedge dz_1+d\bar z_3\wedge dz_2+d\bar z_1\wedge dz_3)\nn
h&=&-\left|c\right|^{2}r^{2}
\eea
and we see that $h$ is constant on the five-sphere.
Another interesting special case is if
one takes $c_2=c_3=c_4=c_5=c_6=0$, since the two-form $W=c_1d\bar z_2\wedge dz_1$
then lives in $\bbR^4\subset \bbR^6$ and the solution preserves twelve supersymmetries. Note that
for this case $h=-|c_1|^2|z_1|^2$ and it vanishes on the locus $z_1=0$.  We can also obtain simple solutions with $W$ real by, for example, taking the real part of
the two-form in \reef{wanst}. To illustrate, a solution with twelve supersymmetries is
obtained if we take $W=c_1d\bar z_2\wedge dz_1+c.c.$ and then $h=-|c_1|^2(|z_1|^2+|z_2|^2)$ which
now vanishes along the lower-dimensional locus $z_1=z_2=0$.

\section{Superisometry algebra}
In this section we will analyse the superisometry algebra for the class of Schr\"odinger invariant solutions discussed
in section 2.4 for a generic $SE_5$, preserving six supersymmetries.

\subsection{Killing vectors}
We begin by presenting the Killing vectors that leave the solution invariant. These correspond to the Hamiltonian $H$,
spatial translations $P_i$,
the number operator $N$, Galilean boosts $G_i$, spatial rotations $M$,
the dilatations $D$,
the special conformal transformations $K$, which together generate the Schr\"odinger algebra,
and the $R$-symmetry of $SE_5$. Explicitly we have:
\begin{align}
H=&\partial_+\nn
P_i=&\partial_i\nn
N=&\partial_-\nn
G_{i}= & -x^{+}\partial_{i}+x^{i}\partial_{-}\nn
M= & x^{2}\partial_{3}-x^{3}\partial_{2}\nn
D= & r\partial_{r}-x^{i}\partial_{i}-2x^{+}\partial_{+}\nn
K= & -2\, x^{+}\, x^{i}\partial_{i}-2\,\left(x^{+}\right)^{2}\,\partial_{+}+\left(x^{i}x^{i}+\frac{1}{r^{2}}\right)\partial_{-}+2\, x^{+}r\, \partial_r\nn
\mathcal{R}= & \partial_{\psi}
\end{align}
where $\partial_\psi$ is the $R$-symmetry Killing vector on $SE_5$ manifold
(see appendix A for more discussion on $SE_5$ spaces). For special choices of $SE_5$ there
could be additional Killing vectors.

Using the ten-dimensional metric, we calculate the dual one-forms, which we will denote by
the same letters hoping that this won't cause any confusion:
\begin{align}
H=&r^2(dx^-+hdx^+)\nn
P_i=&r^2dx^i\nn
N=&r^2dx^+\nn
G_{i}= & r^{2}\left(-x^{+}dx^{i}+x^{i}dx^{+}\right)\nn
{M}= & r^{2}\left(x^{2}dx^{3}-x^{3}dx^{2}\right)\nn
D= & \frac{1}{r}dr-r^{2}x^{i}dx^{i}-2r^{2}\, x^{+}\left(dx^{-}+h\, dx^{+}\right)\nn
{K}= & r^{2}\left(-2\, x^{+}\, x^{i} dx^{i}-2\,\left(x^{+}\right)^{2}\,\left(dx^{-}+h\, dx^{+}\right)+\left(x^{i}x^{i}+\frac{1}{r^{2}}\right)dx^{+}\right)+2\,\frac{x^{+}}{r}\, dr\nn
{\mathcal{R}}= & \eta_{SE}
\end{align}
and $\eta_{SE}$ is the Reeb one-form on the $SE$ manifold.

Actually, it is not immediately obvious that the action of the Reeb Killing vector does in fact leave the solution
invariant for our choice of $W$ and $h$, both of which depend on the coordinates of the $SE_5$ space.
The K\"ahler-form on the $CY_{3}$ cone can be written as
\begin{equation}
\mathcal{J}=r\, dr\wedge\eta+\frac{1}{2}r^{2}\, d\eta_{SE}
\end{equation}
Using this, the (1,1) condition on
$W=d(r^2\sigma)$ then implies that
\begin{align}
d\sigma_{\mu\nu} & =-(\eta_{SE})_{\mu}\left(d\eta_{SE}\right)_{\nu}{}^\rho\sigma_{\rho}+(\eta_{SE})_{\nu}\left(d\eta_{SE}\right)_{\mu}
{} ^\rho\sigma_{\rho}+\frac{1}{4}\left(d\eta_{SE}\right)_{\mu}{}^\rho\left(d\eta_{SE}\right)_{\nu}{}^\sigma
\left(d\sigma\right)_{\rho\sigma}
\end{align}
with indices raised with respect to the metric on $SE_5$.
After using that $|\eta_{SE}|^2=1$ and that $\sigma^\mu$ is a Killing vector on $SE_5$
we deduce that
\begin{align}
\mathcal{L}_{\partial_\psi}\sigma & =0
\end{align}
and it then follows that the Reeb vector still generates a symmetry of the solution.

\subsection{Killing spinor bilinears}
We first observe that if $\ve_1$ and $\ve_2$ are two type IIB Killing spinors then
the ten-dimensional one-form
\be
\bar\ve_1\Gamma_M \ve_2 dx^M +c.c.
\ee
is dual to a Killing vector \cite{HackettJones:2004yi}.
In the following we will calculate such bilinears involving $\ve_P$ and $\ve_S$. In carrying out these calculations
one heavily uses the projection conditions satisfied by $\eta_\pm$. We write
\begin{equation}
\Gamma_{11}=\left(i\Gamma^{+-23}\right)\left(i\Gamma^{4\ldots9}\right)\equiv\gamma_{D3}\gamma_{\left(7\right)}
\end{equation}
and we have
\begin{align}\label{pagain}
\gamma_{D3}\eta_{\pm}= & \pm\eta_{\pm},\qquad
\gamma_{\left(7\right)}\eta_{\pm}=  \eta_{\pm},\qquad\Gamma^+\eta_+=0~.
\end{align}
We also use the conditions arising from $W$ being (1,1) and primitive
\be\label{wpagain}
\sla{W}\eta_\pm=\sla{W}^*\eta_\pm=0~.
\ee

\subsubsection{The $PP$ bilinear}
We define the bilinear form
\be
A\equiv (\bar\ve_P\Gamma_M\ve_P) dx^M
\ee
where here $\Gamma_M$ is a coordinate basis gamma-matrix.
After substituting the expression for $\ve_P$ given in \reef{pf}, and using the projections
\reef{pagain}, \reef{wpagain}
We find
\be
A=\left[r\bar\eta_-\Gamma_M\eta_-
-\frac{r^4}{64}\bar\eta^*_-\Gamma^r\sla{W}^*\Gamma^+\Gamma_M\Gamma^+\sla{W}\Gamma^r\eta^*_-
\right]dx^M~.
\ee
In simplifying the last term, we use
\be
r^2\bar\eta^*_-\Gamma^+\Gamma^r\sla{W}^*\sla{W}\Gamma^r\eta^*_-
=16h\bar\eta_-\Gamma_-\eta_-~.
\ee
A calculation shows that
\be
A=(\bar\eta_-\Gamma_-\eta_-) H+(\bar\eta_-\Gamma_+\eta_-)N+(\bar\eta_-\Gamma_i\eta_- )P^i~.
\ee
This should be compared with
with the equations just below (3.10) in \cite{Sakaguchi:2008ku}.

\def\kin{\eta^K_-}
\def\bkin{\bar\eta^K_-}
\def\dyn{\eta^D_-}
\def\bdyn{\bar\eta^D_-}
We can write
\be
\eta_-=\eta_-^K+\eta_-^D, \qquad \Gamma^+\eta_-^K=0,\qquad \Gamma^-\eta_-^D=0~.
\ee
We then find
\be
A=(\bdyn\Gamma_-\dyn) H+(\bkin\Gamma_+\kin)N+(\bdyn\Gamma_i\kin
+\bkin\Gamma_i\dyn) P^i
\ee
and we see that $\kin$ parametrise the ``kinematical'' supersymmetries found in \cite{dg2}
while $\dyn$ parametrise ``dynamical supersymmetries'' and lead to a positive spectrum for $H$.

\subsubsection{The $PS$ bilinear}
We define the bilinear form
\be
B\equiv  (\bar{\varepsilon}_{P}{\Gamma}_{M}\varepsilon_{S}) dx^{M}+c.c.~.
\ee
After substituting the expressions for $\ve_P$, $\ve_S$ given in \reef{pf}, \reef{sf}, then
using the projection conditions and the primitivity of $W$ one
can show that the only non-zero contribution comes from
\begin{align}
B= & r^{-1}\bar{\eta}_{-}{\Gamma}_{a}f^a_m\Gamma_{r}\eta_{+}\, dx^{m}
-r^2\bar\eta_-\left[\Gamma_+ dx^++\Gamma_i dx^i\right]x^j\Gamma_j\eta_+\nn
&-x^+r^2\bar\eta_-\left[\Gamma_+dx^+\Gamma_i dx^i+\frac{h}{2}\Gamma^+ dx^+\right]\Gamma^-\eta_+
+\frac{r^4}{16}x^+\bar\eta^*_-\Gamma^r\sla{W}^*\sla{W}\Gamma^r\eta_+^* dx^+
+c.c.~.
\end{align}
To proceed we use that
\be
(\bar\eta_-\Gamma_\alpha\Gamma_r\eta_+) \bar f^\alpha=(\bar\eta_-\Gamma_{\alpha=1}\Gamma_r\eta_+) \eta_{SE}
\ee
where we recall that $\bar f^\alpha$ is an orthonormal frame for $SE_5$ and we have taken $\bar f^1\equiv \eta_{SE}$. To see this we use the 4+6 decomposition \reef{etpmdecomp} to write
$\bar\eta_-\Gamma_\alpha\Gamma_r\eta_+=-i\bar q_-q_+\mathcal{J}_{\alpha r}$ with $\mathcal{J}$ given in
\reef{eq:cone_Kahler_form}.
We then find we can express the bilinear form as
\begin{equation}
B=(\bar{\eta}_{-}\eta_{+})\,{D}
+2(\bar{\eta}_{-}\Gamma_{1}\Gamma_{2}\eta_{+})\,{M}
-(\bar{\eta}_{-}\Gamma_{+}\Gamma_{i}\eta_{+})\,{G}_{i}
+(\bar{\eta}_{-}\Gamma_{\alpha=1}\Gamma_{r}\eta_{+})\,{\mathcal{R}}
+c.c. ~.
\end{equation}
This should be compared with the equations just below (3.10) in \cite{Sakaguchi:2008ku}.

\subsubsection{The SS bilinear}

We now consider the bilinear
\begin{align}
C & \equiv(\bar{\varepsilon}_{S}\hat{\Gamma}_{M}\varepsilon_{S}) dx^{M}~.
\end{align}
After substituting the expression for $\ve_S$ given in \reef{sf} it is helpful to observe that, for example,
$\bar\eta_+\Gamma_{A}\eta_+$ can only be non-vanishing if $A=+$.
An easy way to see this is to insert $2=\Gamma^{+}\Gamma_{+}+\Gamma_{+}\Gamma^{+}$.
Using this as well as \reef{pagain}, \reef{wpagain} we see that the only non-zero contribution comes from
the terms
\begin{align}
C= & \Bigg[\frac{1}{r}\,\bar{\eta}_{+}\Gamma_{r}{\Gamma}_{M}\Gamma_{r}\eta_{+}
 -x^{+}\,\left(\bar{\eta}_{+}\Gamma_{r}{\Gamma}_{M}\Gamma_{+}\eta_{+}+\bar{\eta}_{+}\Gamma_{+}{\Gamma}_{M}\Gamma_{r}\eta_{+}
 \right)
  +r\,\bar{\eta}_{+}x^{i}\Gamma_{i}{\Gamma}_{M}x^{j}\Gamma_{j}\eta_{+}\nn
 & +x^{+}r\,\left(\bar{\eta}_{+}x^{a}\Gamma_{a}
 {\Gamma}_{M}\Gamma_{+}\eta_{+}+\bar{\eta}_{+}\Gamma_{+}
 {\Gamma}_{M}x^{a}\Gamma_{a}\eta_{+}\right)
  +\left(x^{+}\right)^{2}\,r\,\bar{\eta}_{+}\Gamma_{+}
 {\Gamma}_{M}\Gamma_{+}\eta_{+}\nn
 &-\frac{r^{3}(x^+)^2}{16}\,\bar{\eta}_{+}^{\ast}\Gamma_{r}\sla{W}^*
 {\Gamma}_{M}\sla{W}\Gamma_{r}\eta_{+}^{\ast}\, \Bigg]dx^{M}~.
 \end{align}
After some further calculation we obtain
\be
C=-(\bar{\eta}_{+}\Gamma_{+}\eta_{+})\hat{K}~.
\ee
This should be compared with the equations just below (3.10) in \cite{Sakaguchi:2008ku}.

\subsection{Generating the superconformal symmetries}
If a supergravity solution has a Killing vector preserving all of the fluxes, then the Lie derivative of
a Killing spinor with respect to that Killing vector generates another Killing spinor. This action
corresponds to the even-odd part of the superisometry algebra.

Here we consider taking the Lie derivative of the Poincar\'e Killing spinors $\ve_P$ with respect to
the special conformal Killing vector $K$. We expect to generate the superconformal Killing spinors.
We have
\begin{equation}
\mathcal{L}_{K}\varepsilon_{P}\equiv K^{M}{\nabla}_{M}\varepsilon_{P}+\frac{1}{8}dK_{MN}{\Gamma}^{MN}\varepsilon_{P}~.
\end{equation}
A calculation reveals that
\begin{align}
\mathcal{L}_{K}\varepsilon_{P}= & \left(-r^{-1}\Gamma_{r}+x^{i}\Gamma_{i} +x^+\Gamma_+\right)\Gamma^{+}\varepsilon_{P}+\frac{rx^{+}}{4}\,\sla{W}\Gamma_{r}\Gamma^{+}\varepsilon_{P}^{\ast}~.\end{align}
This immediately shows that the two Poincare supersymmetries satisfying $\Gamma^{{+}}\varepsilon_P=0$ which
were found in \cite{dg2}, i.e. the two kinematical supersymmetries, cannot generate superconformal Killing spinors.

We now substitute the expression for $\varepsilon_{P}$ to find
\be
\mathcal{L}_{K}\varepsilon_{P}=r^{1/2}\left(-\frac{1}{r}\Gamma_{r}+x^{i}\Gamma_{i}+x^+\Gamma_+\right)
\Gamma^{+}\eta_{-}+\frac{x^{+}r^{3/2}}{4}\,\sla{W}\Gamma_{r}\Gamma^{+}\eta_{-}^{\ast}~.
\ee
and we see that
\be
\mathcal{L}_{K}\varepsilon_{P}=\varepsilon_{S}
\ee
with $\ve_S$ as in \reef{sf} with
\be
\eta_{+}=-\Gamma^{+}\eta_{-}~.
\ee
Thus we see that the special conformal transformations acting on the two extra Poincar\'e Killing spinors, i.e. the dynamical
supersymmetries, generate the two superconformal Killing spinors, as expected.

\section{Conclusion}
In this paper we have carried out a detailed analysis of the supersymmetry that is preserved by a class of solutions
found in \cite{dg2}. We showed that special classes of solutions with Schr\"odinger symmetry can have the supersymmetry
enhanced from two Killing spinors to six, for a generic $SE_5$ space. We also analysed the corresponding superisometry algebra and showed that the two Killing spinors found in \cite{dg2} are kinematical supersymmetries and the four new supersymmetries
consist of two dynamical supersymmetries and two special conformal supersymmetries.
For the special case when
$SE_5$ is the round five-sphere we showed that the supersymmetry can be enhanced from four
Killing spinors to twelve.
For a class of Schr\"odinger$(z)$ invariant solutions found in \cite{dg2} with $z>2$
we showed that while there are no superconformal supersymmetries there can be additional
dynamical supersymmetries. It would be of interest to further extend this analysis to the full
range of supersymmetric solutions with Schr\"odinger($z$) symmetry with $z>2$ found in \cite{dg2}.

It would also be interesting to carry out a similar analysis for the solutions of $D=11$ supergravity with Schr\"odinger$(z)$ symmetry
that were constructed in \cite{dg2}. These solutions share many similarities with the type IIB solutions that we have been considering here and we expect analogous results.

\section*{Acknowledgements}
JPG is supported by an EPSRC Senior Fellowship and a Royal Society Wolfson Award.

\appendix

\section{Some results for $CY_3$ cones}
Consider the cone metric
\be
ds^2(CY_3)=dr^2+r^2ds^2(SE_5)~.
\ee
Using coordinates $x^m=(r,x^\mu)$ we calculate that the non-zero Christoffel symbols are given by
\bea
\Gamma^r_{\mu\nu}&=&-rg^{SE}_{\mu\nu}\nn
\Gamma^\mu_{r\nu}&=&r^{-1}\delta^\mu_\nu\nn
\Gamma^\mu_{\nu\rho}&=&\gamma^\mu_{\nu\rho}
\eea
where $\gamma$ is the Christoffel symbols for $SE_5$.
One can then obtain the result for the Riemann tensor
\be
R^{CY}_{rm np}=0~.
\ee

We next note that if $\sigma$ is an arbitrary one-form on $SE_5$ then it can be pulled back to give a one-form on
$CY_3$. We then have
\bea\label{diffconerw}
\nabla^{CY}_\mu(r^{2}\sigma)_r&=&-r\sigma_\mu\nn
\nabla^{CY}_r(r^{2}\sigma)_\mu&=&r\sigma_\mu\nn
\nabla^{CY}_\mu(r^{2}\sigma)_\nu&=&r^2\nabla^{SE}_\mu\sigma_\nu~.
\eea
In particular, if $\sigma$ is dual to a Killing vector on $SE_5$ then $r^2\sigma$ is dual to a Killing vector on $CY_3$.

Next consider $W=d(r^z \sigma)$ with $\sigma$ a one-form on $SE_5$. We calculate
\bea\label{derivsw}
\nabla^{CY}_r W_{r\mu}&=&z(z-2)r^{z-2}\sigma_\mu\nn
\nabla^{CY}_r W_{\mu\nu}&=&2(z-2)r^{z-1}\nabla^{SE}_{[\mu}\sigma_{\nu]}\nn
\nabla^{CY}_\mu W_{r\nu}&=& r^{z-1}\left[(z-2)\nabla^{SE}_{[\mu}\sigma_{\nu]}+z\nabla^{SE}_{(\mu}\sigma_{\nu)}\right]\nn
\nabla^{CY}_\mu W_{\nu\rho}&=& 2r^z\left[\nabla^{SE}_\mu\nabla^{SE}_{[\nu}\sigma_{\rho]}+zg^{SE}_{\mu[\nu}\sigma_{\rho]}\right]~.
\eea
Note in particular that for the special case when $z=2$ and when the one-form $\sigma$ is dual to a Killing
vector on $SE_5$ (see below), we deduce that
\be\label{z2wd}
\nabla^{CY}_m W_{rn}=0,\qquad \nabla^{CY}_r W_{\mu\nu}=0~.
\ee

If we introduce a frame $f^a=(f^r,f^\alpha)$ on $CY_3$ with $f^r=dr, f^\alpha=r \bar f^\alpha$ where $\bar f^\alpha$ is a frame for $SE_5$
then the covariant derivative of a spinor has coordinate components
\bea
\nabla^{CY}_{m=r}\epsilon &=&\partial_r\epsilon\nn
\nabla^{CY}_{m=\mu}\epsilon &=&\nabla^{SE}_{\mu}\epsilon +\tfrac{1}{2}\bar f^\alpha_\mu\Gamma_\alpha{}^r\epsilon~.
\eea

The metric on $SE_5$ is normalised so that the Ricci tensor is four times the metric.
We will write the metric on $SE_5$ as
\be
ds^2(SE_5)=\eta_{SE}\otimes \eta_{SE} +ds^2(KE_4)
\ee
where $ds^2(KE_4)$ is the transverse K\"ahler-Einstein metric, normalised so that
the Ricci tensor is six times that of the metric, and $d\eta_{SE}=2J_{KE}$ where $J_{KE}$
is the K\"ahler form of $KE_4$. Recall that in general $KE_4$ is only locally defined.
We also write
\be
\eta_{SE}=(d\psi+A), \qquad dA=2J_{KE}
\ee
so that the Reeb Killing vector dual to $\eta_{SE}$ is $\partial_\psi$.
If $\sigma$ is a one-form on $SE_5$ dual to a Killing vector then
\be
\nabla^2_{SE}\sigma_\mu=R^{SE}_\mu{}^\nu\sigma_\nu=-4\sigma_\nu~.
\ee

The K\"ahler form on the cone can be written
as
\bea
\mathcal{J}&=&\nabla^{CY}\left(r^{2}(\eta_{SE})\right)\label{eq:cone_Kahler_form}\nn
&=&r\, dr\wedge\left(\eta_{SE}\right)+r^{2}J_{KE}
\eea
and so in particular
\be\label{jcompexp}
\mathcal{J}_{r\mu}=r(\eta_{SE})_\mu~.
\ee
If $W$ is a $\left(1,1\right)$ from on the cone then
\begin{equation}\label{11pw}
\mathcal{J}_k{}^mW_{ml}=-\mathcal{J}_{l}{}^{m}W_{km}~.
\end{equation}

The $CY_3$ has a covariantly constant, positive chirality spinor $\zeta_+$
and we have
\be\label{defofj}
\mathcal{J}_{mn}=i\zeta_+^\dagger\gamma_{mn}\zeta_+~.
\ee
Furthermore,
\be\label{pcccs}
\gamma_{m}\zeta_{+}=i\mathcal{J}_{m}^{\quad n}\gamma_{n}\zeta_{+}~.
\ee
In holomorphic coordinates\footnote{Note that we use the maths convention that
$\mathcal{J}_{mn}=-I_{mn}$ where $I^m{}_n$ is the complex structure and that in holomorphic
coordinates $I^{i}{}_{j}=i\delta^{i}_{j}$.}
we have $\gamma^{\bar\mu}\zeta_+=0$ and hence
if $W$ is of type (1,1) and primitive and/or type (2,0) (i.e. the (0,2) pieces vanish)
then we have
\be
\tfrac{1}{2}W^*_{mn}\gamma^{mn}\zeta_+=0~.
\ee

Note that if $W=d(r^2\sigma)$ with $\sigma$ an arbitrary one-form on $SE_5$
then $d*_{CY}W=0$ is equivalent to $\nabla_\mu^{SE}\sigma^\mu=0$ and $\nabla_{SE}^2\sigma_\mu=-4\sigma_\mu$.
In turn turn these two conditions are equivalent to $\sigma_\mu$ being dual to a Killing vector on $SE_5$.
If $\sigma_\mu$ is dual to a Killing vector it is simple to see that it implies the two conditions.
Conversely, if we assume the two conditions using an argument in section 4.3 of \cite{pr}
that $\sigma_\mu$ is dual to a Killing vector.

\section{Solving equation \eqref{theheqn}}
We would like to solve
\begin{align}
\left(\dslash h -\frac{r}{8}\sla{W}\sla{W}^*\gamma^r\right)\zeta_{+} =0
\end{align}
on the $CY_3$ cone subject to $W$ being  (1,1) and primitive i.e. satisfying
$\sla{W}\zeta_+=0$, $\sla{W}^*\zeta_+=0$. We first recast the condition in the form
\begin{align}
\gamma^n\left( \partial_n h -\frac{r}{2}W_{nm}W^{*m}{}_r\right)\zeta_{+} =0~.
\end{align}
To proceed we now use the projection condition on the covariantly constant spinor \reef{pcccs}
to obtain
\begin{equation}
\gamma^{n}\left[\partial_nh -\frac{r}{4}\left(W_{nm}W^{*m}{}_r
+W_{nm}^{\ast}W^m{}_{r}\right)
+i\frac{r}{4}\mathcal{J}_{n}{}^{k}\left(W_{km}W^{*m}{}_r-  W_{km}^{\ast}W^m{}_r\right)\right]\zeta_{+}=0\end{equation}
where $\mathcal{J}_{mn}$ is the K\"ahler-form on $CY_3$. This expression is of the form $\gamma^nT_n\zeta_+$ with $T_n$ real.
After multiplying by $\gamma^m T_m$ we conclude that $T_n=0$:
\be
\partial_nh=
-\partial_n\left(|S|_{CY}^2\right)      -i\frac{r}{4}\mathcal{J}_{n}{}^{k}\left(W_{km}W^{*m}{}_r-  W_{km}^{\ast}W^m{}_r\right)~.
\ee
Here we have introduced the one-form $S$ that is dual to a Killing vector on the CY cone defined by
\be
S\equiv r^2\sigma~.
\ee
We can now solve this for $h$:
\be
h=-\left(|S|^2_{CY}-i\mathcal J^{mn}S_mS^*_n\right)~.
\ee
This can be verified using $\nabla^{CY}_mS_n=(1/2)W_{mn}$ and also \reef{11pw}.
This expression for $h$ is actually negative definite. This can be seen by writing it in the form
\be
h=-2|S^{(0,1)}|_{CY}^2
\ee
where
\be
S^{(0,1)}_m\equiv\frac{1}{2}\left(S_m+i\mathcal{J}_m{}^n S_n\right)~.
\ee

We can express $h$ in yet another way by first introducing a one-form $\lambda$ on the $SE_5$
space given by
\be
\lambda_\mu\equiv i(\mathcal{L}_{\sigma^{\ast}}^{SE_{5}}\sigma_\mu)\equiv i(\sigma^{*\nu}\nabla^{SE}_\nu\sigma_\mu
-\sigma^{\nu}\nabla^{SE}_\nu\sigma^*_\mu)\ee
where the notation means that we are taking the Lie-derivative with respect to the vector field
which is dual, with respect to the $SE_5$ metric, to $\sigma^*$. Next, using \reef{11pw}
and also \reef{jcompexp} we deduce that we can write $h$ as
\begin{equation}
h=-r^{2}\left(\left|\sigma\right|_{SE}^{2}+\tfrac{1}{2}(\eta_{SE})^{\mu}\lambda_{\mu}\right),
\label{eq:h_solution_A}~.
\end{equation}

Finally we also need to ensure that the equation of motion $\nabla^{2}_{CY}h+|W|^2_{CY}=0$ in
\reef{eqsom}, arising from Einstein's equations, is satisfied. We find that this is equivalent to
\begin{equation}
\nabla^2_{SE_{5}}\left((\eta_{SE})^{\mu}\lambda_{\mu}\right)=-12\, (\eta_{SE})^{\mu}\lambda_{\mu}\end{equation}
Given that $\eta$ and $\lambda$ are one-forms on $SE_5$ that are dual to Killing vectors, this condition
is equivalent to demanding that the two form on $CY_3$ given by
\begin{equation}
V=dL
\end{equation}
is primitive where
\be
L\equiv r^2\lambda
\ee
is dual to a Killing vector on $CY_3$.
We can also show that $V$ is a (1,1) form on $CY_3$.
We have
\be
W_{mn}=2\nabla^{CY}_mS_n,\qquad V_{mn}=2\nabla^{CY}_m L_n \label{W_def}
\ee
and it is straightforward to show that
\be
L=i\mathcal{L}^{CY}_{S^*} S~.
\ee
Combining these results we calculate that
\begin{equation}\label{benas}
V_{mn}=\frac{i}{2} \left(W^{\ast}_{mk}W^{k}{}_{n}-W^{\ast}_{nk}W^{k}{}_{m}\right)+i R_{mnkl}S^{k}S^{*l}
\end{equation}
where for the second term we used the result that for any Killing vector
\begin{equation}
\nabla_{m}\nabla_{n}S_{k}=-R_{nkml}S^{l}
\end{equation}
and also the Bianchi identity for the Riemann tensor. The term in the brackets in \reef{benas}
is $(1,1)$, since $W$ is (1,1), and so is the second term since $R$ is the Riemann tensor of a K\"ahler metric.

Note that if $S^{(0,1)}=0$ then $h=0$ and from the equation of motion
$\nabla^{2}_{CY}h+|W|^2_{CY}=0$ we see that $W=\sigma=0$.

\section{Killing spinors for $z>2$}
For $W=d(r^z\sigma)$ with $z>2$ the analysis of the Killing spinor equations
proceeds in exactly the same way as in section 2 up to equation \reef{pcds}.
We next substitute $m=r$ into \reef{covp1}, \reef{covp2} and use \reef{projpms} to deduce that
\be
\Gamma^+\nabla^{CY}_r\sla{W}\Gamma^r\e^*_+=0~.
\ee
From \reef{pcds} we also have
\be
\Gamma^+\nabla^{CY}_r\sla{W}\e^*_+=0~.
\ee
Together these imply $\Gamma^+\nabla^{CY}_r W_{r\mu}\Gamma^\mu\epsilon_+^*=0$ and hence, after using \reef{derivsw}, that for $z\ne 2$
\be
\Gamma^+(\sigma_\alpha\Gamma^\alpha)\epsilon^*_+=0~.
\ee
Combining this with \reef{sigone}, we deduce that for $\sigma\ne 0$ we necessarily have
\be
\Gamma^+\epsilon_+=0~.
\ee

The remaining equations that one finds are very similar to the $z=2$ case.
Let us label $\epsilon_+=\psi_+$. We find that $\psi_+$ has only dependence on
the $CY_3$ coordinates and $\nabla^{CY}_m\psi_+=0$. Furthermore,
$\sla{W}^*\psi_+=\sla{W}\psi_+=0$
and so if $\psi_+\ne 0$ then $W$ is (1,1) and primitive. We also find
\bea
\Gamma^+\sla{W}\e_-=\Gamma^+\sla{W}^*\e_-=0\label{c1}\\
\nabla^{CY}_m\epsilon_-+\frac{1}{8}\Gamma^+\sla{W}\Gamma_m\e_-^*=0\label{c2}\\
\Gamma^+\dslash{h}\e_-+\sla{W}\e_-^*=0\label{c3}\\
\e_-=\psi_--x^+(\Gamma^-\psi_++\frac{r}{4}\sla{W}\Gamma^r\psi_+^*)=0~.\label{c4}
\eea
Substituting \reef{c4} into \reef{c2} we obtain
\bea
\nabla^{CY}_m\psi_-+\frac{1}{8}\Gamma^+\sla{W}\Gamma_m\psi_-^*=0\label{surely}\\
\nabla^{CY}_m W^{\ast}_{rn}\Gamma^{n}\psi_+=0~.
\eea
From the second equation we obtain the two constraints
\begin{align}
\left(z-2\right)\sigma_{\alpha}^{\ast}\Gamma^{\alpha}\psi_{+} & =0\label{lake1}\\
\left[\left(z-1\right)\nabla^{SE}_{\alpha}\sigma_{\beta}^{\ast}+\nabla^{SE}_{\beta}
\sigma_{\alpha}^{\ast}\right]\Gamma^{\beta}\psi_{+}&=0~.\label{lake2}\end{align}

\subsection{No superconformal Killing spinors for $z>2$}
We now show that $\psi_+=0$. Let us assume the converse and then
$\psi_{+}$ is a covariantly constant spinor on $\mathbb{R}^{1,3}\times CY_{3}$.
Using the 4+6 decomposition \reef{tendgam}
we can write
\begin{equation}
\psi_{+}=q_{+}\otimes\zeta_{+}~.
\end{equation}
With $q_{+}\ne 0$, equations \reef{lake1}, \reef{lake2} become
\begin{align}
\sigma_{\alpha}^{\ast}\gamma^{\alpha}\zeta_{+} & =0\label{eq:zg2_1}\\
\left[\left(z-1\right)\nabla^{SE}_{\alpha}\sigma_{\beta}^{\ast}+\nabla^{SE}_{\beta}\sigma_{\alpha}^{\ast}\right]\gamma^{\beta}\zeta_{+} & =0~.
\label{eq:zg2_2}\end{align}
Multiplying equation \eqref{eq:zg2_1} by $\zeta^{\dagger}\gamma_{r}$
from the left we obtain (using \reef{defofj} and \reef{eq:cone_Kahler_form})
\begin{equation}
\sigma_{1}=0\end{equation}
where we are using an orthonormal frame $\bar f^\alpha$ on $SE_5$ with $\bar f^1=\eta_{SE}$.
On the other hand multiplying equation \eqref{eq:zg2_2} by $\zeta^{\dagger}\gamma_{r}$
from the left we have
\begin{align}
d\sigma_{1 \alpha}^{\ast}+z\nabla^{SE}_{\alpha}\sigma_{1}^{\ast} & =0\\
\Rightarrow d\sigma_{1 \alpha}^{\ast} & =z(J_{KE})_{\alpha}{}^{\beta}\sigma_{\beta}^{\ast}\label{getclose}\\
\end{align}
where we used that the spin connection on $SE_5$ has components $\bar\omega^1{}_\alpha=(J_{KE})_{\alpha\beta}\bar f^\beta$.
We now use the $\left(1,1\right)$ condition on $W$ to deduce
\begin{align}
W_{1 \alpha} & =\mathcal{J}_{1}{}^{r}\mathcal{J}_{\alpha}{}^{n}W_{rn}\\
\Rightarrow d\sigma_{1 \alpha} & =-z\, (J_{KE}){\alpha}{}^{\beta}\sigma_{\beta}
\label{eq:dbeta2}\end{align}
which in combination with \eqref{getclose} implies $\sigma=0$ and hence for $z>2$:
\begin{equation}
\psi_{+}=0.
\end{equation}

\subsection{Example for $z>2$}
Despite the fact that $\psi_+=0$ we can have special solutions
that have enhanced supersymmetry with $\Gamma^+\psi_-\ne0$. Let us illustrate
for the special case when $SE_5$ is the round five-sphere.

We construct a closed, primitive (1,1) form $W$ using a $\left(0,1\right)$ one-form $A(z)$ on $\bbR^6$:
\begin{equation}
W=dA=\partial_{{z}^{i}}A_{\bar{z}_{i}}\left({z}\right)\, d{z}^{i}\wedge d\bar{z}^{j}
\end{equation}
(implicitly we are assuming that it takes the form $W=d(r^z\sigma$)) with the prmitivity condition
fixed by choosing
that $A$ is co-closed
\begin{equation}
\partial_{{z}^{i}}A^{{z}^{i}}=0~.
\end{equation}
This also guarantees that $d*_{CY}W=0$.

Equation \reef{surely}
reads
\begin{equation}
\partial_{m}\psi_{-}-\frac{1}{4}\Gamma^{+}W_{mn}\Gamma^{n}\psi_{-}^{\ast}=0~.\label{eq:Ex1_spin_eq}\end{equation}
From \reef{c2} we observe that $\nabla^{CY}_m(\Gamma^{+}\psi_{-})=0$
and we restrict our attention to Killing spinors satisfying
\begin{equation}
\Gamma^{\bar{z}^{i}}\Gamma^{+}\psi_{-}=0~.
\end{equation}
This brings equation \eqref{eq:Ex1_spin_eq} to the form\begin{align}
\partial_{m}\psi_{-}-\frac{1}{4}\Gamma^{+}\nabla_{m}\,\sla A\psi_{-}^{\ast} & =0
\end{align}
which we can solve as follows:
\begin{align}
\psi_{-}=  \eta_{-}+\frac{1}{4}\Gamma^{+}\sla A\,\eta_{-}^{\ast},\qquad
\partial_{m}\eta_{-}=  0~.\end{align}
Next, equation \eqref{c3} fixes
\begin{equation}
h=-\tfrac{1}{2}\left|A\right|^{2}\end{equation}
One can check that the equation of motion
$\nabla^{2}_{CY}h+|W|^2_{CY}=0$ is also satisfied.
The Killing spinors take the form
\begin{equation}
\varepsilon=r^{\frac{1}{2}}\,\left(\eta_{-}+\tfrac{i}{4}\Gamma^{+}\sla A\,\eta_{-}^{\ast}\right)~.
\end{equation}

\end{document}